\documentclass[pra,aps,twocolumn,nopacs,superscriptaddress,nofootinbib,longbibliography,notitlepage]{revtex4-1}

\usepackage{amsmath}  \usepackage{amssymb}  \usepackage{amsfonts}  \usepackage{mathrsfs}  \usepackage{mathtools} \usepackage{bm}  \usepackage{bbm}   \usepackage{braket}  \usepackage{color}  \usepackage{comment}  \usepackage{dcolumn}  \usepackage{enumerate}  \usepackage{epsfig}  \usepackage{gensymb}  \usepackage{graphicx}  \usepackage{indentfirst}  \usepackage{lmodern}   \usepackage{psfrag}  \usepackage{pst-all}  \usepackage{soul}  
\usepackage{xcolor}
\usepackage{float} 
\usepackage[colorlinks,linkcolor=blue,citecolor=blue,urlcolor=blue,hyperindex,driverfallback=dvipdfm]{hyperref}  \usepackage[T1]{fontenc} 

\def\ii{{\rm i}}  \def\ee{{\rm e}}
  
\def\Ree{{\rm Re}}  \def\Imm{{\rm Im}}
        \def\Eb{{\bf E}}                      \def\kb{{\bf k}}        \def\pb{{\bf p}}      \def\Rb{{\bf R}}  \def\rb{{\bf r}}      \def\vb{{\bf v}} 
\def\xx{\hat{\bf x}}  \def\yy{\hat{\bf y}}  \def\zz{\hat{\bf z}}        \def\eh{\hat{\bf e}}  \def\RR{\hat{\bf R}}  
\def\kpar{k_\parallel}  \def\kparb{{\bf k}_\parallel} 
\def\EF{{E_{\rm F}}}     
\def\lamp{\lambda_{\rm p}}    \def\kp{k_{\rm p}}    \def\wbulk{\omega_{\rm bulk}} 
  \def\red{\color{red}}    
\def\p{{\rm p}}  \def\s{{\rm s}}  \def\d{{\rm d}}    \def\elec{{\rm e}}    \def\wD{\omega_{\rm D}}  \def\kD{k_{\rm D}}

\begin{document} 
\renewcommand{\thesection}{\arabic{section}} 
\renewcommand{\thesubsection}{\arabic{subsection}} 
\def\bibsection{\section*{\refname}} 

\title{Free-electron coupling to surface polaritons mediated by small scatterers
}

\author{Leila~Prelat}
\author{Eduardo~J.~C.~Dias}
\affiliation{ICFO-Institut de Ciencies Fotoniques, The Barcelona Institute of Science and Technology, 08860 Castelldefels (Barcelona), Spain}
\author{F.~Javier~Garc\'{\i}a~de~Abajo}
\email{javier.garciadeabajo@nanophotonics.es}
\affiliation{ICFO-Institut de Ciencies Fotoniques, The Barcelona Institute of Science and Technology, 08860 Castelldefels (Barcelona), Spain}
\affiliation{ICREA-Instituci\'o Catalana de Recerca i Estudis Avan\c{c}ats, Passeig Llu\'{\i}s Companys 23, 08010 Barcelona, Spain}


\begin{abstract}
The ability of surface polaritons (SPs) to enhance and manipulate light fields down to deep-subwavelength length scales enables applications in optical sensing and nonlinear optics at the nanoscale. However, the wavelength mismatch between light and SPs prevents direct optical excitation of surface-bound modes, thereby limiting the widespread development of SP-based photonics. Free electrons are a natural choice to directly excite strongly confined SPs because they can supply field components of high momentum at designated positions with subnanometer precision. Here, we theoretically explore free-electron--SP coupling mediated by small scatterers and show that low-energy electrons can efficiently excite surface modes with a maximum probability reached at an optimum surface--scatterer distance. By aligning the electron beam with a periodic array of scatterers placed near a polariton-supporting interface, in-plane Smith--Purcell emission results in the excitation of surface modes along well-defined directions. Our results support using scattering elements to excite SPs with low-energy electrons.
\end{abstract}

\maketitle 
\date{\today} 

\section{Introduction}

Surface polaritons (SPs) are collective charge oscillations occurring at the interfaces of a wide range of materials and offering a platform to control light--matter interactions at the nanoscale \cite{S11,paper283}. Specifically, surface-bound optical modes allow us to concentrate electromagnetic energy down to deep-subwavelength length scales and produce large enhancements of the associated near electric fields \cite{paper156}. Polaritons in two-dimensional (2D) materials are particularly promising because they are highly confined \cite{DFM14,paper335} and can be widely controlled by external stimuli introduced through electrical gating \cite{KYC05,JGH11,YSZ13,FRA12}, chemical doping \cite{LLZ11,BYC21}, magnetic fields \cite{WGH13}, or optical heating \cite{NWG16,JKW16,paper337,paper345}. These appealing properties enable technological applications in areas such as biosensing \cite{AHL08,B12,paper308,RMF21}, photodetection \cite{YVF06,KMA14,paper346}, light harvesting \cite{AP10,TYC18}, nonlinear optics \cite{PN08,KZ12,PSL18,paper337}, and optoelectronics \cite{BSH10,paper235,ZWR20}.

Unfortunately, strong field confinement in SPs comes at a price, as these modes cannot be directly excited by far-field radiation due to the kinematical mismatch that maintains them trapped to the surface. This problem limits practical applications of SPs, so much effort has been devoted to tackling it through different approaches, including the use of prisms \cite{KR1968} and nanotips \cite{HKK01,DSB20}. However, prisms are impractical when the SP in-plane wavelength $\lamp$ is small compared with the light wavelength $\lambda_0=2\pi c/\omega$ at the same frequency $\omega$ because of the lack of materials with sufficiently high refractive index $>\lambda_0/\lamp\gg1$. In addition, coupling elements such as gratings \cite{CB10} modify the SP characteristics and offer poor spatial precision, while small coupling nanostructures such as tips \cite{ANG14} are inefficient \cite{paper331} unless strong conditions are met including a precise angular profile of the external light \cite{paper370}.

As an alternative, electron beams (e-beams) combine high spatial precision with an efficient coupling to SPs \cite{paper149}. These probes were instrumental in pioneering studies of surface plasmons \cite{PS1959,PSV1975} and are currently reaching a simultaneous spectral/spatial resolution in the few-meV/sub-nm range using electron energy-loss spectroscopy \cite{KLD14,LTH17,HNY18,LB18,HKR19,HHP19,HRK20,YLG21} (EELS). The detection of cathodoluminescence light emission resulting from the e-beam interaction with a specimen at a nanometer-controlled position also permits studying bright modes with a spectral resolution depending on the optical spectrometer and the signal-to-noise ratio \cite{SMS19,paper338}. Recently, within the emerging field of ultrafast electron microscopy, a temporal resolution in the sub-fs regime \cite{FES15,YFR21,paper415} has become a possibility by teaming up pulsed lasers and free electrons, while also enabling electron spectromicroscopy to be performed with sub-nm/sub-meV resolution \cite{H99,paper114}. It should be noted that e-beams effectively act as broadband nanoscale optical sources and, therefore, lack mode selectivity. In particular, an electron moving with velocity $v$ parallel to a polariton-supporting planar surface can excite a wide spectrum of modes satisfying the condition $\omega<\kpar v$, where $\omega$ and $\kpar$ are the mode frequency and in-plane wave vector, respectively. In this configuration, the emission is delocalized along the electron interaction path, and relativistic electrons are needed to excite SPs whose dispersion lies close to the light cone. These limitations could be potentially addressed by combining electron excitation with small structures placed in the vicinity of the surface and acting as coupling elements.

In this article, we theoretically explore the excitation of SPs in planar surfaces by parallel e-beams assisted by small scatterers. Specifically, we obtain a clean SP signal by operating under conditions in which electrons cannot directly excite surface modes (i.e., for $\omega>\kpar v$), so that the scatterers act as intermediate elements where a dipole is induced containing the high-momentum components that are needed to excite SPs. With this approach, narrow-band SP emission can be realized if the scatterer features a spectrally localized mode. We demonstrate that the SP excitation efficiency can be maximized when the scatterer is placed at an optimum distance from the surface, as illustrated by studying the generation of surface-plasmon polaritons (SPPs) in a thin film (e.g., graphene and atomically thin metal layers). Perfectly resonant scatterers and hBN disks are considered. The latter constitute a practical realization of deep-subwavelength resonant scatterers operating in the mid-infrared spectral region. We further extend the phenomenon of Smith--Purcell emission \cite{SP1953} from photons to polaritons by placing a periodic particle array above a polariton-supporting surface and exciting it by a parallel e-beam, leading to the generation of collimated SPs at angles depending on the periodicity, the electron velocity, and the mode dispersion relation. Like in the individual particle, an optimum scatterer--surface distance is also obtained for polaritonic Smith--Purcell emission. These results pave the way for using free electrons combined with small scatterers to control the generation of polaritons in planar surfaces.


\begin{figure*}[htbp]
\centering
\includegraphics[width=0.8\textwidth]{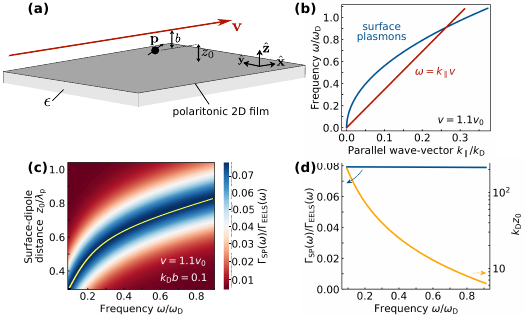}
\caption{\textbf{Free-electron excitation of surface polaritons (SPs) mediated by a dipolar scatterer}. \textbf{(a)}~Geometry under consideration, consisting of an electron moving with velocity $\vb$ parallel to a polariton-supporting surface (a film supported on a substrate of real permittivity $\epsilon$) and passing above a small particle placed at distances $b$ from the trajectory and $z_0$ from the surface. The electron induces a dipole $\pb$ on the particle, which couples to the SPs. The particle is in the plane defined by $\vb$ and the surface normal. \textbf{(b)}~Dispersion relation of surface-plasmon polaritons (SPPs) $\omega/\wD=\sqrt{\kpar/\kD}$ for a 2D film described by a Drude conductivity of weight $\wD$ [Eq.~(\ref{Drude})] with $\kD=(\epsilon+1)\hbar\wD/4\pi e^2$ (e.g., $\wD=\EF/\pi\hbar$ for graphene) compared with the electron line $\omega=\kpar v$. Both curves intersect at $\omega/\wD=v_0/v$, where $v_0=\wD/\kD\approx0.092\,c/(\epsilon+1)$. We take $v=1.1\,v_0$. \textbf{(c)}~Spectrally resolved probability of SPP emission $\Gamma_{\rm SP}(\omega)$ under the conditions of (a,b) as a function of excitation frequency $\omega$ and particle--surface separation $z_0$ for fixed $\kD b=0.1$ (e.g., $\approx9.5$~nm for self-standing graphene with $\EF=0.3$~eV). We assume a lossless resonant scatterer and normalize $\Gamma_{\rm SP}(\omega)$ to the EELS probability $\Gamma_{\rm EELS}(\omega)$ in the absence of the 2D layer [Eq.~(\ref{eq_Gamma_EELS})], $\omega$ to $\wD$, and $z_0$ to the SPP wavelength $\lamp$. \textbf{(d)}~Maximum $\Gamma_{\rm SP}(\omega)$ along the yellow curve in (c) (left vertical scale) and normalized optimum separation $\kD z_0$ (right scale) as a function of frequency.}
\label{Fig1} 
\end{figure*}

\section{Electron excitation of surface polaritons mediated by a dipolar particle}
\label{sec_1_dipole}

We consider the interaction between a fast electron and an individual dipolar scatterer placed at the origin, a distance $z_0$ away from a planar surface lying at the $z=-z_0$ plane, as schematically sketched in Figure~\ref{Fig1}(a). The electron moves parallel to the surface with a velocity $\vb=v\xx$ at a distance $b$ above the scatterer [i.e., the electron trajectory is given by $\rb_\elec(t)=(v t,0,b)$]. We describe the optical response of the scatterer through a frequency-dependent diagonal polarizability tensor $\alpha(\omega)=\alpha_\parallel(\omega)\RR\otimes\RR+\alpha_\perp(\omega)\zz\otimes\zz$ [adopting the notation $\Rb=(x,y)$], while the surface is introduced through Fresnel's reflection coefficients. Using these elements and starting from the field of the electron in free space, we calculate the surface-reflected field and the self-consistent dipole induced at the scatterer. The result is then analyzed in far in-plane regions to quantify the number of polaritons excited by the passage of the electron.

The electron can directly excite SPs (without the presence of a scatterer) under the condition that the line $\omega=\kpar v$ lies above the mode dispersion curve in $\kpar-\omega$ space [i.e., for energies above the crossing point in Figure~\ref{Fig1}(b)]. While the interplay between direct and particle-assisted SP creation constitutes an interesting scenario for future analysis, we limit the present work to frequencies $\omega$ below the SP excitation threshold (i.e., direct SP excitation by the electron is kinematically forbidden, but SPs can still be created through the mediation of particle polarization).

\subsection{Electron field near a planar surface}

Working in frequency space $\omega$, the electric field directly created by the electron as it moves in free space can be written as \cite{J99}
\begin{align}
\Eb_\elec^{\rm dir}(\rb,\omega)=\frac{\ii e}{v} \int_{-\infty}^{\infty} &\frac{\d k_y}{\kappa} \,  \ee^{\ii\kparb\cdot\Rb-\kappa|z-b|} \label{eq1}\\
&\times\Big(\frac{\omega}{v\gamma^2},k_y,\ii\,\kappa\,{\rm sign}\{z-b\}\Big), \nonumber
\end{align}
where $\kparb=(\omega/v)\xx+k_y\yy$ is the in-plane wave vector, $\kappa=\sqrt{(\omega/v\gamma)^2+k_y^2}$ describes an evanescent decay away from $z=b$, and $\gamma=1/\sqrt{1-v^2/c^2}$. This equation shows that the electron field is composed of components satisfying the condition $\kpar\le\omega/v$, corroborating that modes above the electron line in Figure~\ref{Fig1}(b) cannot be directly excited by the electron.

The presence of the planar surface at $z=-z_0$ produces a reflected field that can be conveniently expressed in terms of Fresnel's reflection coefficients $r_{\kpar\p}$ and $r_{\kpar\s}$ for p and s polarization, respectively. To do so, we note that the integrand in Eq.~(\ref{eq1}) is orthogonal to the wave vector $\kparb+\ii\,{\rm sign}\{z-b\}\kappa\zz$ (i.e., the field is transverse), so it can be projected on p- and s-polarization components represented by the unit vectors $\eh_{\kparb\p}^\pm= (1/k\kpar)(\pm k_z\kparb- \kpar^2\zz)$ (with $k=\omega/c$, $k_z=\sqrt{k^2-\kpar^2+\ii0^+}$, and ${\rm Im}\{k_z\}>0$) and $\eh_{\kparb\s}^\pm= (1/k_\parallel)(-k_y\xx + k_x\yy)$, respectively, which we eventually evaluate at $k_x=\omega/v$ and $k_z=\ii\kappa$. Here, upper (lower) signs correspond to waves propagating upward (downward). The direct electron field is then expressed in the form
\begin{subequations}
\begin{align}
\Eb_\elec^{\rm dir}(\rb,\omega)=\frac{ek}{v} \int_{-\infty}^{\infty} &\frac{\d k_y}{\kpar} \,  \ee^{\ii\kparb\cdot\Rb-\kappa(b-z)} \label{eq_Ee_dir}\\
&\times\Big[-\eh_{\kparb\p}^-+\frac{\ii k_yv}{\kappa\,c}\,\eh_{\kparb\s}^-\Big] \nonumber
\end{align}
for $z<b$ (i.e., it consists of downward waves emanating from the e-beam). By applying the reflection coefficients to each wave vector and polarization component, the reflected field in the $z>-z_0$ region above the surface is found to be
\begin{align}
\Eb_\elec^{\rm ref}(\rb,\omega)=\frac{ek}{v} \int_{-\infty}^{\infty} &\frac{\d k_y}{\kpar} \,  \ee^{\ii\kparb\cdot\Rb-\kappa(z+b+2z_0)} \label{eq_Ee_ref}\\
&\times\Big[-r_{\kpar\p}\eh_{\kparb\p}^++r_{\kpar\s}\frac{\ii k_yv}{\kappa\,c}\,\eh_{\kparb\s}^+\Big]. \nonumber
\end{align}
\end{subequations}
The total field generated by the electron in the presence of the surface then becomes $\Eb_\elec(\rb,\omega)=\Eb_\elec^{\rm dir}(\rb,\omega)+\Eb_\elec^{\rm ref}(\rb,\omega)$. Here, we assume $b>0$ (i.e., the electron moves above the scatterer). An extension to $b<0$ (electron moving in the region separating the surface from the scatterer) is straightforward but should not add qualitatively different results.

\subsection{Self-consistent particle-near-surface response}
\label{sec_self_response}

The field generated by the electron polarizes the dipolar particle placed at the origin, which displays a dipole moment \begin{align}
\pb(\omega)=\alpha(\omega)\cdot\Eb(0,\omega) \label{palphaE}
\end{align}
in the frequency domain. Here, $\Eb(0,\omega) = \Eb_\elec(0,\omega) + \Eb_{\rm dip}^{\rm ref}(0,\omega)$ is the total field acting on the particle, which consists of the electron field and the reflection of the dipole field by the surface $\Eb_{\rm dip}^{\rm ref}(0,\omega)$.

We follow a similar procedure as with the electron field to express $\Eb_{\rm dip}^{\rm ref}(0,\omega)$ in terms of p- and s-polarization components, starting from the direct dipole field \cite{J99} $\Eb_{\rm dip}^{\rm dir}(\rb,\omega)=[k^2\pb(\omega)+\pb(\omega)\cdot\nabla\,\nabla]\ee^{\ii kr}/r$ and projecting it on in-plane wave-vector and polarization components to yield \cite{paper370}
\begin{subequations}
\label{dipolefields}
\begin{align}
\Eb_{\rm dip}^{\rm dir}(\rb,\omega)=\frac{\ii k^2}{2\pi} \int &\frac{\d^2\kparb}{k_z} \, \ee^{\ii \kparb\cdot\Rb+\ii k_z|z|} \\
&\times \sum_{\sigma={\rm s,p}} \big[\pb(\omega)\cdot\eh_{\kparb\sigma}^s\big]\eh_{\kparb\sigma}^s, \nonumber
\end{align}
where $s={\rm sign}\{z\}$. The surface-reflected field is then obtained by introducing the reflection coefficients as
\begin{align}
\Eb_{\rm dip}^{\rm ref}(\rb,\omega)=\frac{\ii k^2}{2\pi} \int &\frac{\d^2\kparb}{k_z} \, \ee^{\ii \kparb\cdot\Rb+\ii k_z(z+2z_0)} \label{diprefall}\\
&\times \sum_{\sigma={\rm s,p}} r_{\kpar\sigma}\,\big[\pb(\omega)\cdot\eh_{\kparb\sigma}^-\big]\eh_{\kparb\sigma}^+. \nonumber
\end{align}
\end{subequations}
Finally, setting $\rb=0$ (the dipole position) and carrying out the azimuthal integral in $\kparb$, we find
\begin{align}
\Eb_{\rm dip}^{\rm ref}(0,\omega)=\mathcal{G}(\omega)\cdot\pb(\omega)
\label{dipref}
\end{align}
in terms of a Green tensor $\mathcal{G}(\omega)=\mathcal{G}_\parallel(\omega)\RR\otimes\RR+\mathcal{G}_\perp(\omega)\zz\otimes\zz$ with in- and out-of-plane components given by
\begin{align}
\left[\begin{array}{c} \mathcal{G}_\parallel(\omega) \\ \mathcal{G}_\perp(\omega) \end{array} \right]
=\frac{\ii}{2}\int_0^\infty\frac{\kpar\d\kpar}{k_z}\,\ee^{2\ii k_zz_0}\,
\left[\begin{array}{c} k^2\,r_{\kpar{\rm s}}-k_z^2\,r_{\kpar{\rm p}} \\ 2\kpar^2\,r_{\kpar{\rm p}} \end{array} \right].
\nonumber
\end{align}
Combining this result with Eq.~(\ref{palphaE}), we can write the self-consistently induced dipole as
\begin{subequations}
\label{pandalphaeff}
\begin{align}
\pb(\omega)=\alpha^{\rm eff}(\omega)\cdot\Eb_\elec(0,\omega), \label{palphaeff}
\end{align}
where
\begin{align}
\alpha^{\rm eff}(\omega)=\frac{1}{1/\alpha(\omega)-\mathcal{G}(\omega)} \label{alphaeff}
\end{align}
\end{subequations}
is an effective polarizability that takes into account the effect of the surface, so only the electron field is now required to evaluate the induced dipole [cf. Eqs.~(\ref{palphaE}) and (\ref{palphaeff})].

\subsection{The limit of strongly confined surface polaritons}
\label{limitofstrongconfinement}

We are interested in generating SPs of small wavelength $\lamp$ compared with the light wavelength $\lambda_0=2\pi c/\omega$. In this limit, we can neglect the contribution of s-polarization components to the surface response (i.e., $r_{\kpar\s}\to0$) and approximate $k_z\approx\ii\kpar$ in the dipolar self-interaction, so the Green tensor reduces to
\begin{align}
\mathcal{G}_\perp(\omega)\approx2\,\mathcal{G}_\parallel(\omega)\approx\int_0^\infty \kpar^2\d\kpar\,\ee^{-2\kpar z_0}\,r_{\kpar{\rm p}}. \label{Gqs}
\end{align}
For simplicity, we take the e-beam to intersect the surface normal at the particle position. Again, we neglect s-polarization components in the surface-reflected electron field, but retain the relativistic $\gamma$ factors, so that our results remain valid at high electron velocities, even though the response of the particle--surface system is treated electrostatically. Then, evaluating Eq.~(\ref{eq_Ee_ref}) at the particle position ($\rb=0$), we find
\begin{subequations}
\label{ErefandEdir}
\begin{align}
\Eb_\elec^{\rm ref}(0,\omega)=-\frac{\ii e}{v} \int_{-\infty}^{\infty}\d k_y &\,  \ee^{-\kappa(b+2z_0)} \label{Eerefqs}\\
&\times r_{\kpar\p}\,(\kappa\omega/v\kpar^2,0,\ii).
\nonumber
\end{align}
For the direct electron field, we can analytically integrate Eq.~(\ref{eq1}), which yields \cite{paper149}
\begin{align}
\Eb_\elec^{\rm dir}(0,\omega)=\frac{2e\omega}{v^2\gamma}\bigg[\frac{\ii}{\gamma}K_0\Big(\frac{\omega b}{v\gamma}\Big)\xx+K_1\Big(\frac{\omega b}{v\gamma}\Big)\zz\bigg],
\label{Edirana}
\end{align}
\end{subequations}
where $K_m$ are modified Bessel functions.

The $\lamp\ll\lambda_0$ condition is generally satisfied if the surface consists of a thin film, which, neglecting nonlocal and retardation effects, can be described through a frequency-dependent 2D surface conductivity $\sigma(\omega)$ and treated as a zero-thickness layer. When the film is deposited on a substrate of real permittivity $\epsilon$, the Fresnel reflection coefficient for p polarization from the vacuum side reduces to \cite{GP16}
\begin{align}
\label{eq_pole_approx}
r_{\kpar\p}= 1+\mathcal{R}_\p\dfrac{\kp}{\kpar-\kp},
\end{align}
where $\mathcal{R}_{\rm p}=1/\bar{\epsilon}$ with $\bar{\epsilon}=(\epsilon+1)/2$, while $\kp=\ii\omega\bar{\epsilon}/2\pi\sigma(\omega)$ is the complex polariton wave vector identified as a pole in $r_{\kpar\p}$ \cite{JBS09,paper235}, whose existence requires fulfilling the condition ${\rm Im}\{\sigma\}>0$. The SP wavelength is then defined as $\lamp=2\pi/{\rm Re}\{\kp\}$.

We consider long-lived polaritons characterized by low inelastic losses (i.e., ${\rm Re}\{\sigma\}\ll|\sigma|$) such as, for example, SPPs in high-quality doped graphene \cite{GPN12,paper235,NMS18}, whose surface conductivity can be approximated as
\begin{align}
\sigma(\omega)=\frac{e^2}{\hbar}\frac{\ii\wD}{\omega+\ii\gamma} \label{Drude}
\end{align}
in the Drude model, where $\wD=\EF/\pi\hbar$ is a frequency weight proportional to the doping Fermi energy $\EF$ and we introduce a phenomenological inelastic damping rate $\gamma$. This type of response is also found in thin noble metal films, where $\wD=\hbar\wbulk^2 d/4\pi e^2$ depends on the bulk plasma energy $\hbar\wbulk$ (e.g., $9.17$~eV in silver \cite{JC1972}) and the film thickness $d$ \cite{paper335}. Assuming the surface conductivity in Eq.~(\ref{Drude}), the SPP dispersion relation obtained from the $\kpar=\kp$ pole in Eq.~(\ref{eq_pole_approx}) is universally given by $\kpar=(\bar{\epsilon}\hbar/2\pi e^2\wD)\,\omega(\omega+\ii\gamma)$ and exhibits a characteristic $\omega\propto\sqrt{\kpar}$ scaling, as shown in Figure~\ref{Fig1}(b).

Using the reflection coefficient in Eq.~(\ref{eq_pole_approx}), the $\kpar$ integral in Eq.~(\ref{Gqs}) can be performed analytically, leading to
\begin{align}
\mathcal{G}_\perp(\omega)&=2\,\mathcal{G}_\parallel(\omega)
=\frac{1}{4z_0^3}
+\mathcal{R}_\p\kp^3\Big[\frac{1}{\theta}+\frac{1}{\theta^2}-\ee^{-\theta}\,{\rm Ei}(\theta)\Big] \nonumber
\end{align}
with $\theta=2\kp z_0$ and Ei denoting the exponential integral function (see Eq.~3.353-5 of Ref.~\citenum{GR1980}). The $\omega$ dependence of the Green tensor is then encapsulated in $\kp$. In the low-loss limit ($\gamma\ll\omega$), we can approximate
\begin{align}
{\rm Im}\{r_{\kpar\p}\}\approx\pi\mathcal{R}_\p\kp\delta(\kpar-\kp)
\label{Imrpll}
\end{align}
with $\kp=2\pi/\lamp$ taken as a real number. Inserting this expression into Eq.~(\ref{Gqs}), we find
\begin{align}
{\rm Im}\{\mathcal{G}_\perp(\omega)\}=2\,\mathcal{G}_\parallel(\omega)=\pi\mathcal{R}_\p\kp^3\ee^{-2\kp z_0}. \label{ImGG}
\end{align}
We use this result in what follows.

In the low-loss limit, noticing that the polariton pole dominates the surface response, we can also obtain an analytical expression for the surface-reflected electron field by calculating the residue due to the $\kpar=\kp$ pole of $r_{\kpar\p}$ in Eq.~(\ref{Eerefqs}). To this end, we multiply and divide the $\mathcal{R}_\p$ term in the integrand by $\kpar-\kp$, close the integration contour in the complex $k_y$ plane, and ignore the contributions of branching points and any other poles in the integrand. The field contributed by the first term in the right-hand side of Eq.~(\ref{eq_pole_approx}) takes the same form as the direct field in Eq.~(\ref{Edirana}), but with $b$ replaced by $b+2z_0$ and the sign of the $x$ component flipped. In this polariton-pole approximation (PPA), inserting the result in Eq.~(\ref{palphaeff}) together with the direct field from Eq.~(\ref{Edirana}), the electron field in the absence of the particle reduces to
\begin{align}
\Eb_\elec(0,\omega)&\approx
\dfrac{2e\omega}{v^2} \label{Eefinal}\\
\times\bigg[
&\bigg\{
 \frac{\ii}{\gamma^2}K_0\Big[\frac{\omega b}{v\gamma}\Big]
-\frac{\ii}{\gamma^2}K_0\Big[\frac{\omega(b+2z_0)}{v\gamma}\Big] \nonumber\\
&\quad\quad\quad\quad\quad\quad\quad
+\dfrac{\pi\mathcal{R}_\p\nu\,\ee^{-\nu\omega(b+2z_0)/v}}{\sqrt{\mu^2-1}}\bigg\}\xx, \nonumber\\
+&\bigg\{
 \dfrac{1}{\gamma}K_1\Big[\frac{\omega b}{v\gamma}\Big]
+\dfrac{1}{\gamma}K_1\Big[\frac{\omega(b+2z_0)}{v\gamma}\Big] \nonumber\\
&\quad\quad\quad\quad\quad\quad\quad
+\dfrac{\ii\pi\mathcal{R}_\p\mu^2\ee^{-\nu\omega(b+2z_0)/v}}{\sqrt{\mu^2-1}}\bigg\}\zz \nonumber
\bigg],
\end{align}
where $\mu=\kp v/\omega$ and $\nu=\sqrt{\mu^2-v^2/c^2}$. The first, second, and third terms inside each of the curly brackets in Eq.~(\ref{Eefinal}) origine from the direct electron field [Eq.~(\ref{Edirana})], the surface-reflected field contributed by the unity term in $r_{\kpar\p}$, and the result of the remaining part of $r_{\kpar\p}$, respectively.

\subsection{Polariton excitation probability}
\label{gammaw}

Under the conditions of Figure~\ref{Fig1}(a,b), assuming that the electron line $\omega=\kpar v$ lies below the mode dispersion curve, direct SP excitation by the electron is kinematically forbidden, so SPs are exclusively excited {\it via} the induced particle dipole [Eq.~(\ref{palphaeff})]. We now calculate the number of the so-emitted surface quanta. To this end, we first consider a dipole $\pb(t)$ placed in free space. By integrating the radial Poynting vector over both time and the surface of a large sphere centered at the dipole, we find a total emitted energy $(2/3\pi c^3)\int_0^\infty\omega^4\d\omega\,|\pb(\omega)|^2$. We then divide each frequency component by $\hbar\omega$ to obtain the total number of emitted quanta $\int_0^\infty\d\omega\,\Gamma_0(\omega)$, where $\Gamma_0(\omega)=(2\omega^3/3\pi\hbar c^3)|\pb(\omega)|^2$ gives the spectral decomposition of the emission probability. This result differs by a factor of $1/2\pi$ from the decay rate of a transition dipole in free space \cite{paper053}.

For clarity, we note that the time-domain induced dipole is obtained from Eq.~(\ref{palphaeff}) by performing the Fourier transform $\pb(t)=(2\pi)^{-1}\int\d\omega\,\ee^{-\ii\omega t}\pb(\omega)$. Now, following the same analysis as in Ref.~\citenum{paper053}, we find that the presence of a material structure changes the spectral probability to $\Gamma(\omega)=\Gamma_0(\omega)+(1/\pi\hbar)\,{\rm Im}\{\pb^*(\omega)\cdot\Eb^{\rm ind}\}$, where $\Eb^{\rm ind}$ is the self-induced electric field at the dipole position. In particular, near a planar surface, this field is given by $\Eb^{\rm surf}_{\rm dip}(0,\omega)$ in Eq.~(\ref{dipref}), and consequently, $\Gamma(\omega)=\Gamma_0(\omega)+(1/\pi\hbar)\,{\rm Im}\{|\pb_\parallel(\omega)|^2\mathcal{G}_\parallel(\omega)+|p_z(\omega)|^2\mathcal{G}_\perp(\omega)\}$. Adopting again the electrostatic limit [Eq.~(\ref{Gqs})], this leads to
\begin{align}
\Gamma_{\rm SP}(\omega)=\frac{1}{\pi\hbar}\,\big[|\pb_\parallel(\omega)|^2+2|p_z(\omega)|^2\big]\,{\rm Im}\{\mathcal{G}_\parallel(\omega)\}
\label{GSP}
\end{align}
for the spectrally resolved probability of emitting SPs, where ${\rm Im}\{\mathcal{G}_\parallel(\omega)\}$ is proportional to $\mathcal{R}_\p$, indicating that $\Gamma_{\rm SP}(\omega)$ is only contributed by SP emission within the geometry and approximations under consideration. Incidentally, this result coincides with the integral of the absorption density within the 2D film (see Sec.~\ref{polempro} in Methods).

We aim to maximize the probability of exciting surface modes, which is enhanced when $\alpha^{\rm eff}(\omega)$ increases [see Eqs.~(\ref{palphaeff})], and therefore, optimum emission should occur if the scatterer is lossless (i.e., composed of nonabsorbing materials, such that $\Imm\{-1/\alpha_s(\omega)\}=2k^3/3$ for each of the polarization directions $s=\parallel,\perp$) and resonant (i.e., $\Ree\{1/\alpha_s(\omega)-\mathcal{G}_s(\omega)\}=0$, a condition that can be generally fulfilled at specific resonance frequencies \cite{paper370}). Combining these two conditions, we have
\begin{align}
\alpha^{\rm eff}_s(\omega)=\frac{\ii}{2k^3/3+\Imm\{\mathcal{G}_s(\omega)\}} \label{alphalosslessresonant}
\end{align}
for the effective polarizability of a perfect scatterer.

To illustrate the magnitude of $\Gamma_{\rm SP}(\omega)$ in Eq.~(\ref{GSP}) and its dependence on geometrical parameters, we find it convenient to normalize it to the EELS probability produced by a free-standing particle (without the surface) \cite{paper371} $\Gamma_{\rm EELS}(\omega)=(4e^2\omega^2/\pi\hbar v^4\gamma^2)$ $\big[(1/\gamma^2)K^2_0\left(\omega b/v\gamma\right)\Imm\{\alpha_\parallel\}$ $+K_1^2\left(\omega b/v\gamma\right)\Imm\{\alpha_\perp\}\big]$, which we evaluate for a perfect scatterer by setting $\alpha(\omega)=3\ii/2k^3$:
\begin{align}
\Gamma_{\rm EELS}(\omega)=\dfrac{6\,e^2c^3}{\pi\hbar v^4\gamma^2\omega}
\bigg[\frac{1}{\gamma^2}K^2_0\Big(\frac{\omega b}{v\gamma}\Big)
+ K_1^2\Big(\frac{\omega b}{v\gamma}\Big)\bigg].  \label{eq_Gamma_EELS}
\end{align}
The resulting ratio $\Gamma_{\rm SP}(\omega)/\Gamma_{\rm EELS}(\omega)$ is presented in Figure~\ref{Fig1}(c) as a function of frequency and particle--surface separation $z_0$ for a fixed impact parameter $b$,
assuming a lossless resonant particle [Eq.~({\ref{alphalosslessresonant})] and a 2D Drude material described by the conductivity of Eq.~({\ref{Drude}).

Despite the indirect mechanism for SP generation, the probability plotted in Figure~\ref{Fig1}(c) takes sizeable values about one order of magnitude lower than the maximum possible EELS probability for a dipolar scatterer. This figure also reveals the presence of a maximum in the emission probability $\Gamma_{\rm SP}(\omega)$ for an optimum value of $z_0$. The latter depends on SP frequency and lies in the $z_0<\lamp$ range [Figure~\ref{Fig1}(c), yellow line]. As a function of SP frequency, this maximum remains nearly constant over the explored frequency range [Figure~\ref{Fig1}(d)] for the particular choice of the chosen parameters. We note that Figure~\ref{Fig1}(d,b) is calculated analytically using Eqs.~(\ref{palphaeff}), (\ref{ImGG}), (\ref{GSP}), and (\ref{alphalosslessresonant}) combined with the PPA expression for the electron field in Eq.~(\ref{Eefinal}), but almost identical results are obtained when the latter is computed without approximations from the sum of Eqs.~(\ref{ErefandEdir}) (Supplementary Figure~\ref{FigS1}).

Upon inspection of the analytical expressions in Eqs.~(\ref{ImGG}) and (\ref{Eefinal}), we find that ${\rm Im}\{\mathcal{G}(\omega)\}$ and the PPA electron field $\Eb_\elec(0,\omega)$ depend on $\omega$ and $z_0$ only through the ratios $\omega/\wD$ and $z_0/\lamp$, and consequently, the PPA approach yields universal results for a lossless resonant scatterer and a thin film described through a low-loss Drude conductivity [Eq.~(\ref{Drude})]. The SP emission probability is also dependent on damping and impact parameter through $\gamma/\wD$ and $b/\lamp$. To corroborate this result, we show that the emission probabilities obtained for graphene and a thin silver film (also described in the Drude model as explained in Sec.~\ref{appendix optical response}; see Methods) yield nearly identical results as a function of $\omega/\wD$ and $z_0/\lamp$ (Supplementary Figure~\ref{FigS2}).

\subsection{Excitation of graphene plasmons assisted by a hBN disk}

So far, we have considered lossless, resonant scatterers to mediate the excitation of SPPs in Drude-like films. We now show that similar results are obtained when these ideal conditions are relaxed and we consider resonant hBN particles to excite mid-infrared plasmons in graphene. These particles have low losses, although still much higher than those associated with radiative emission in free space (see Supplementary Figure~\ref{FigS4}). However, losses arising from SP emission when the particles are placed close to the surface become dominant at short separations, so the small intrinsic damping due to material losses in the particle does not play a major effect.

\begin{figure*}[htbp]
\centering
\includegraphics[width=1.0\linewidth]{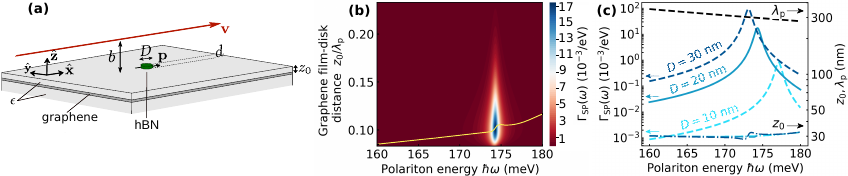}
\caption{\textbf{Free-electron excitation of graphene plasmons mediated by a hBN disk scatterer}. \textbf{(a)}~Geometry under consideration, comprising a graphene film and an hBN disk of diameter $D$ and thickness $d$. The disk is directly deposited on the top dielectric layer of thickness $z_0$. The electron passes at a distance $b$ above the disk. The dielectric substrate and coating layer have a permittivity $\epsilon$. \textbf{(b)}~Spectrally resolved probability of graphene-plasmon emission $\Gamma_{\rm SP}(\omega)$ under the conditions of (a) as a function of plasmon energy $\hbar\omega$ and coating-layer thickness $z_0$ for $D=20$~nm, $d=1$~nm, $\epsilon=2$, $b=10$~nm, and $v=0.04\,c$. We set the graphene Fermi energy to $\EF=1$~eV. \textbf{(c)}~Maximum $\Gamma_{\rm SP}(\omega)$ (left vertical scale, solid curve) and optimum coating thickness $z_0$ along with the SP wavelength $\lamp$ (right scale) as a function of emission frequency under the conditions of (b). In (c), we also show results for the maximum $\Gamma_{\rm SP}(\omega)$ with $D=10$~nm and 30~nm (see labels).}
\label{Fig2}
\end{figure*}

We then consider a feasible configuration in which a hBN disk (thickness $d$ and diameter $D\gg d$) acts as a relatively lossless scatterer that couples to SPs in a graphene film. To make the system more realistic, we assume the graphene film to be supported on a dielectric substrate (permittivity $\epsilon$) and coated by a dielectric film of thickness $z_0$ (same material). The hBN disk is taken to be directly deposited on the surface of the coating layer, with vacuum above it [Figure~\ref{Fig2}(a)]. Again, we assume that all the involved distances are small compared with the light wavelength, so the system can be described in the electrostatic limit. In addition, the thickness of the hBN disk is assumed to be small compared with the polariton wavelength, and consequently, it can be represented by a surface conductivity (see Sec.~\ref{appendix optical response} in Methods).

The effective disk polarizability cannot be calculated by following the approach used above for a self-standing scatterer because the Green tensor diverges as the disk--surface separation is reduced. Instead, we use an analytical expression for the polarizability of a disk deposited on a homogeneous substrate obtained from an electrostatic modal expansion \cite{paper303}, and then introduce the effect of the graphene layer through an {\it ad hoc} Green tensor, as shown in Methods (Sec.~\ref{appendix polarizability}). In addition, the surface reflection coefficient needs to be modified to account for the new layered structure. A Fabry-Perot-type of analysis readily leads to
\begin{align}
r_{\kpar\p}=\frac{r_0+\ee^{-2\kpar z_0}}{1+r_0r'_{\kpar\p}\ee^{-2\kpar z_0}},  \label{tilderp}
\end{align}
where $r_0=(\epsilon-1)/(\epsilon+1)$ and $r'_{\kpar\p}=\kpar/(\kpar-\kp')$ with $\kp'=\ii\omega\epsilon/2\pi\sigma(\omega)$ are the reflection coefficients of the homogeneous $\epsilon$ surface and the $\epsilon$-embedded graphene layer \cite{GP16}, respectively.

Finally, the SP emission probability differs for the geometry in Figure~\ref{Fig2}(a) relative to Figure~\ref{Fig1}(a). A calculation based on the absorption density within the graphene film yields the result (see Sec.~\ref{polempro} in Methods)
\begin{align}
\Gamma_{\rm SP}(\omega)=\frac{1}{\pi\hbar}\,|p_x(\omega)|^2\,{\rm Im}\{\mathcal{G}_\parallel(\omega)\},
\label{GSPhBN}
\end{align}
similar to Eq.~(\ref{GSP}) but with $p_z=0$ (i.e., the induced disk dipole is parallel to the surface) and a modified Green tensor
\begin{align}
{\rm Im}\{\mathcal{G}_\parallel(\omega)\}&\approx\Big(\frac{2\epsilon}{\epsilon+1}\Big)^2 \label{GqshBN}\\
&\times\int_0^\infty\kpar^2\d\kpar\,\ee^{-2\kpar z_0}\,\frac{{\rm Im}\{r'_{\kpar{\rm p}}\}}{\big|1+r_0r'_{\kpar\p}\ee^{-2\kpar z_0}\big|^2}, \nonumber
\end{align}
which reduces to Eq.~(\ref{Gqs}) when setting $\epsilon=1$.

We now use Eq.~\eqref{GSPhBN} combined with Eq.~(\ref{GqshBN}) to calculate the surface-polariton emission probability in the configuration of Figure~\ref{Fig2}(a). Here, the induced dipole $p_x(\omega)=\alpha^{\rm eff}_\parallel(\omega)E_{\elec,x}(0,\omega)$ is obtained by using the modified polarizability in Eqs.~(\ref{alphadisk}) (see Sec.~\ref{appendix polarizability} in Methods) together with the electron field given by the sum of Eqs.~(\ref{ErefandEdir}). In particular, Eq.~(\ref{Eerefqs}) is evaluated by inserting the multilayer reflection coefficient given in Eq.~(\ref{tilderp}) with $z_0=0$ because the electron moves at a distance $b$ from the outer surface on which the direct electron field is reflected.

In Figure~\ref{Fig2}(b), we plot the resulting probability $\Gamma_{\rm SP}(\omega)$ as a function of disk--2D-layer spacer $z_0$ and frequency around the upper hBN Reststrahlen band for a disk thickness $d=1$~nm (corresponding to 3 monolayers) and diameter $D=20$~nm$\gg d$. This choice for the disk diameter is motivated to enable a dipolar resonance within that band and boost the polarizability. In agreement with Figure~\ref{Fig1}, coupling to graphene plasmons is also maximized for an optimum frequency-dependent value of $z_0/\lamp$. However, we now observe a strong dependence of $\Gamma_{\rm SP}(\omega)$ on mode frequency, with a sharp increase at a specific spectral position that is controlled by the intrinsic diameter-dependent resonance of the disk [see Figure~\ref{Fig2}(b) for $D=20$~nm and Figure~\ref{Fig2}(c) for $D=10$, 10, and 30~nm; see also Supplementary Figure~\ref{FigS4}(a) for the disk resonance frequency]. This feature demonstrates that the coupling scheme here considered allows us to select the excitation frequency of narrowband plasmons in a region where they would otherwise be inaccessible via direct electron excitation. Interestingly, the optimum $z_0/\lamp$ ratio takes smaller values than in Figure~\ref{Fig1}, presumably as a result of the weak scattering strength of the hBN disk compared to a perfect scatterer (see Supplementary Figure~\ref{FigS4}). This hypothesis is consistent with the observation that the maximum $\Gamma_{\rm SP}(\omega)$ and the optimum $z_0$ are both increasing with disk diameter [see the probability curves for $D=10$, 20, and 30~nm in Figure~\ref{Fig2}(c)].

\begin{figure*}[htbp]
\centering
\includegraphics[width=\linewidth]{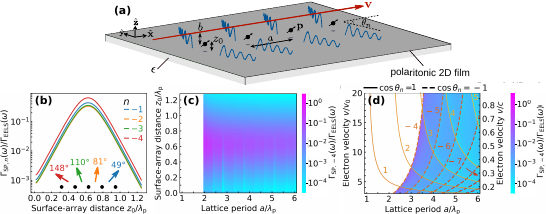}
\caption{\textbf{Polaritonic Smith--Purcell effect}. \textbf{(a)}~Scheme with the geometry under consideration, similar to Figure~\ref{Fig1}(a) but with a linear array of particles spaced by a period $a$ along the direction of electron motion. The in-plane emission angle $\theta_n$ is indicated for one of the orders $n$ [see Eq.~(\ref{eq_angle_SP})]. \textbf{(b)}~Probability of SPP emission $\Gamma_{{\rm SP},n}(\omega)$ for different orders $n$ normalized per particle and divided by the self-standing-particle EELS probability $\Gamma_{\rm EELS}(\omega)$ [Eq.~(\ref{eq_Gamma_EELS})] as a function of scaled array--surface distance $z_0/\lamp$ for fixed frequency $\omega=0.4\,\wD$. We assume perfect scatterers, a Drude film like that in Figure~\ref{Fig1}, and parameters $a/\lamp=2$, $v=2.2\,v_0\approx0.2\,c/(\epsilon+1)$, and $\kD b=0.1$. The corresponding emission angles $\theta_n$ are indicated in the inset (top-view scheme). \textbf{(c,d)}~Ratio $\Gamma_{{\rm SP},n=-4}(\omega)/\Gamma_{\rm EELS}(\omega)$ as a function of either (c) $a/\lamp$ and $z_0/\lamp$ for $v/v_0=2.2$, or (d) $a/\lamp$ and $v/v_0$ for $z_0/\lamp=0.6$, with all other parameters the same as in (b). The right vertical scale in (d) shows $v/c$ for $\epsilon=1$. White regions in (c,d) correspond to the condition that the $n=-4$ Smith--Purcell emission is kinematically forbidden. Probability minima contours are observed as light-blue regions in (d) and marked by solid and dashed curves corresponding to the conditions $\cos\theta_n=1$ and $-1$ respectively, for the values of $n$ indicated by labels.}
\label{Fig3}
\end{figure*}

\section{Polaritonic Smith--Purcell emission}
\label{sec_inf_dipoles}

Smith--Purcell radiation emission \cite{SP1953} occurs when an e-beam is oriented parallel to a grating such that there is a coherent superposition in the far-field emission associated with periodic elements. A "simple Huygens construction" \cite{SP1953} shows that broadband light is emitted along wavelength-dependent directions satisfying the phase-matching relation $\cos\theta_n=n\lambda_0/a+c/v$, where $a$ is the period and $\theta_n$ is the emission angle relative to the electron velocity vector $\vb=v\xx$ when consecutive grooves contribute with a relative time delay given by an integral number $n$ of optical periods \cite{paper149}. Here, we extend this concept to the emission of polaritons by generalizing the results of Sec.~\ref{sec_1_dipole} to a linear array of scatterers, as depicted in Figure~\ref{Fig3}(a). From the Huygens construction sketched in the figure, we find that the SP emission produced by an electron running parallel to the array is subject to the modified phase-matching condition
\begin{align}
\label{eq_angle_SP}
& \cos\theta_n = \lambda_{\p} \left(\dfrac{ n }{a}+\dfrac{c}{v \lambda_0}\right).
\end{align}
Incidentally, while $n<0$ is required by the traditional version of this effect to obtain a real emission angle $\theta_n$ for free photons, polaritonic emission with a positive $n$ is also possible for electrons moving faster than the SP phase velocity (i.e., $v>c\,\lamp/\lambda_0$) and $\lamp<a$. Here, we consider parameters for which direct excitation of polaritons (coincident with $n=0$ emission) is kinematically forbidden.

\subsection{Electron interaction with a linear array of scatterers}

To analyze polaritonic Smith--Purcell emission, we assume again low inelastic losses, so that SPs can propagate far from the interaction region, where the direct and surface-reflected electron fields are negligible at frequencies below the crossing of the electron line and the mode dispersion [see Figure~\ref{Fig1}(b)]. In those far surface regions, the polariton field is dominated by the contribution of the $\kpar=\kp$ pole in $r_{\kpar\p}$ [Eq.~(\ref{eq_pole_approx})] to $\Eb_{\rm dip}^{\rm ref}(\rb,\omega)$ [Eq.~(\ref{diprefall})]. In the electrostatic limit, this contribution to the field reduces to $\Eb_{\rm dip}^{\rm ref}(\rb,\omega)=-\nabla\phi_{\rm dip}^{\rm ref}(\rb,\omega)$, written in terms of the associated scalar potential
\begin{align}
\phi_{\rm dip}^{\rm ref}(\rb,\omega)&=\int \frac{\d^2\kparb}{(2\pi)^2} \, \ee^{\ii \kparb\cdot\Rb}
\,\tilde{\phi}_{\rm dip}^{\rm ref}(\kparb,z,\omega) \label{onedipole}
\end{align}
with
\begin{align}
\tilde{\phi}_{\rm dip}^{\rm ref}(\kparb,z,\omega)&=2\pi \,\ee^{-\kpar(z+2z_0)} \,r_{\kpar\sigma}
\,\pb(\omega)\cdot(\ii\hat{\kb}_\parallel+\zz) \nonumber
\end{align}
for an individual dipole placed at the origin.

We now consider an infinite linear array of period $a$ formed by dipolar scatterers at positions $\Rb_j=ja\xx$, indexed by an integer number $j$. Because of periodicity, the external electron potential at a generic scatterer $j\neq0$ differs from the potential at scatterer $j=0$ by just a phase factor $\ee^{\ii(\omega/v)aj}$ [see Eq.~(\ref{eq1})], and this dependence is directly imprinted on the self-consistent dipole fields. Therefore, the potential produced by all dipoles in the array can be written as $\phi_{\rm array}^{\rm ref}(\rb,\omega)=\sum_{j=-\infty}^\infty\ee^{\ii(\omega/v)aj}\phi_{\rm dip}^{\rm ref}(\rb-\Rb_j,\omega)$ in terms of the potential of one dipole at $j=0$ [Eq.~(\ref{onedipole})]. We can perform the $j$ sum by using the identity $\sum_{j=-\infty}^\infty\ee^{\ii(\omega/v-k_x)aj}=(2\pi/a)\sum_{n=-\infty}^{\infty}\delta(k_x-k_{nx})$, where $k_{nx}=\omega/v+2\pi n/a$ with $n$ running over diffraction orders. The $\delta$-functions in this expression allow us to readily carry out the $k_x$ integral and find the result
\begin{align}
\phi_{\rm array}^{\rm ref}(\rb,\omega)&=\frac{1}{2\pi a}\sum_{n=-\infty}^\infty\int \d k_y \, \ee^{\ii\kb_{n\parallel}\cdot\Rb}
\,\tilde{\phi}_{\rm dip}^{\rm ref}(\kb_{n\parallel},z,\omega) \label{phiarray1}
\end{align}
with $\kb_{n\parallel}=k_{nx}\xx+k_y\yy$.

At large transverse distances $|y|$ from the array, we expect the potential to be dominated by SP components emanating from the $\kpar=\kp$ pole in Eq.~(\ref{eq_pole_approx}). Using the same PPA procedure that allowed us to perform the $k_y$ integral in Eq.~(\ref{Eerefqs}) and obtain Eq.~(\ref{Eefinal}), we transform Eq.~(\ref{phiarray1}) into
\begin{align}
\phi_{\rm array}^{\rm ref}(\rb,\omega)\approx&\frac{2\pi\ii\,\mathcal{R}_\p\kp^2}{a}
\,\pb(\omega)\cdot(\ii\hat{\kb}_{n\parallel}+\zz) \label{phireffinal}\\
&\times\sum_{n=-\infty}^\infty \frac{1}{k_{ny}} \,\ee^{\ii\kb_{n\parallel}\cdot\Rb-\kp(z+2z_0)}, \nonumber
\end{align}
with $k_y\to k_{ny}=\sqrt{\kp^2-k_{nx}^2}$. The far in-plane field is then found to be made of different diffracted SP plane waves labeled by $n$. The associated wave vectors $\kb_{n\parallel}$ form angles $\theta_n$ with the $x$ axis (i.e., the e-beam) determined by $\cos\theta_n=k_{nx}/\kp$. This condition can directly be recast into Eq.~(\ref{eq_angle_SP}) by neglecting the imaginary part of the SP wave vector and writing $\kp=2\pi/\lamp$.

Periodicity also allows us to write the induced dipoles as $\pb(\omega)\,\ee^{\ii(\omega/v)aj}$, where the dependence on the position of each particle $j$ reduces to a phase factor. In the absence of interaction among particles, $\pb(\omega)$ would be given by Eqs.~(\ref{pandalphaeff}). However, particle--particle interaction may become relevant for strong scatterers and small separations, so we include it in our calculations by defining the effective polarizability of each particle in the array as in Eq.~(\ref{alphaeff}) but with the Green tensor supplemented by the contribution of the direct and surface-reflected fields produced by the rest of the dipoles. This is calculated in Methods (see Sec.~\ref{latticesum}), and in what follows, we use Eq.~(\ref{alphaeff}) combined with Eq.~(\ref{latticesumfinal}) to obtain $\alpha^{\rm eff}(\omega)$ for particles in the array.

To obtain the number of SPs produced by the passage of the electron, we consider the power density transported by a mode characterized by an electrostatic electric field $E_0(\xx+\ii\zz)\ee^{\kpar(\ii x-z-z_0)-\ii\omega t}+{\rm c.c.}$ in the vacuum region outside ($z>-z_0$) a polariton-supporting planar surface placed at the $z=-z_0$ plane. The power per unit of length along the transverse direction $y$ is given by \cite{paper331} $I_\p=|\omega E_0^2/(2\pi\mathcal{R}_\p k_\p^2)|$, where we assume a reflection coefficient $r_{\kpar\p}\approx\mathcal{R}_\p\kp/(\kpar-\kp)$ dominated by a pole at $\kpar=\kp$ and neglect inelastic losses (i.e., ${\rm Im}\{\kp\}\ll|\kp|$, so $\kp$ is treated as a real number).
We now substitute $E_0$ by the in-plane field amplitude associated with the $n$-th diffracted beam in Eq.~(\ref{phireffinal}) [i.e., $2\pi|\mathcal{R}_\p\kp^2\,(\ii p_x\cos\theta_n+p_z)|/a\sin\theta_n$, where we have used the relation $k_{ny}=\kp\sin\theta_n$], then divide by $\hbar\omega$ to transform energy into number of SP quanta, and finally multiply by the length $a\sin\theta_n$ (the period projected on the SP wavefront direction). By following this procedure, we obtain the expression
\begin{align}
\Gamma_{{\rm SP},n}(\omega)=\frac{8\pi^3|\mathcal{R}_\p|\,|\ii\, p_x(\omega)\cos\theta_n+p_z(\omega)|^2}{\hbar\, \lamp^2\,a\sin\theta_n} \label{Gammaarray}
\end{align}
for the number of emitted SPs normalized to the number of scatterers in the array. For a given set of geometrical parameters and frequency, Smith--Purcell emission is only allowed for orders $n$ satisfying the condition $-\kp\leq 2\pi n/a+\omega/v\leq\kp$. In addition, Eq.~(\ref{Gammaarray}) suggests that the Smith--Purcell emission at a given order $n$ is boosted when polaritons are emitted along the direction of the array (i.e., $\sin\theta_n=0$), as we corroborate below.

We consider an array of perfect scatterers {like the one sketched in Figure~\ref{Fig3}(a)} and evaluate $\Gamma_{\text{SP},n}(\omega)$ using Eq.~(\ref{Gammaarray}) with $\pb(\omega)$ calculated from Eqs.~(\ref{palphaeff}), (\ref{Eefinal}), and (\ref{alphalosslessresonant}). The latter involves $s=xx$ and $zz$ components evaluated from Eq.~(\ref{latticesumfinal}) (notice that the direct lattice sum does not contribute to the imaginary part of the Green tensor). An illustrative result is presented in Figure~\ref{Fig3}(b) for all allowed emission orders $n$, normalized to the perfect-scatterer EELS probability [Eq.~(\ref{eq_Gamma_EELS})]. Like in the configuration studied in Figure~\ref{Fig1} for an individual scatterer, we find again a maximum emission at an optimum array--surface separation $z_0$, which is roughly independent of the order $n$. In Figure~\ref{Fig3}(c), we show the dependence of $\Gamma_{{\rm SP},n}$ for the strongest allowed mode $n=-4$ as a function of lattice period $a$ and surface--array distance $z_0$. We observe singularities in the probability at specific values of $a$ corresponding to the onset of new emission orders $n<-1$ (see Supplementary Figure~\ref{FigS5}). Nevertheless, the optimum value of $z_0$ that maximizes the emission is roughly independent of $a$. Incidentally, $n=-4$ emission is kinematically forbidden in the white regions in Figure~\ref{Fig3}(c,d). For each emission order $n$, plasmonic Smith--Purcell emission can thus be finely tuned by playing with the electron velocity, the period, and the array--surface separation.

Interestingly, similar singularities of $\Gamma_{{\rm SP},{-4}}$ as a function of period $a$ and electron velocity $v$ are present in Figure~\ref{Fig3}(d), corresponding to light-blue contours in the color plot. In particular, contours with a positive slope emerge from the condition $\theta_n=\pi$ (i.e., anti-parallel emission relative to the electron velocity vector) for different orders $n=-4,-5,\cdots$ (from left to right). The origin of these features is analogous to lattice resonances in particle arrays \cite{paper090}, emerging when the wavelength matches the lattice period, or equivalently, at the onset of every diffraction order. Ultimately, this is the result of the accumulation of in-phase fields emanating from the different particle dipoles in the array \cite{R1907}. In our system, the array period is small compared with the light wavelength, so free light propagation cannot produce in-phase contributions. However, fields propagating as surface modes can be in phase when the polariton wavelength $\lamp$ is close to the period $a$. This translates into a divergent-like behavior of the lattice sum in Eq.~(\ref{Gsurffinal}) (see Supplementary Figure~\ref{FigS5} for a plot of this sum), that in turn permeates the induced dipoles and the resulting emission probability in Eq.~(\ref{Gammaarray}). Likewise, features in Figure~\ref{Fig3}(d) emerging as negative-slope contours also originate in lattice resonances for forward-propagating polaritons at $\theta_n=0$ with $n=1,2,\cdots$ from left to right.

\begin{figure*}[htbp]
\centering
\includegraphics[width=\linewidth]{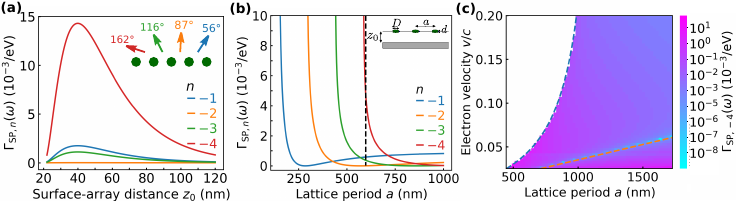}
\caption{\textbf{Smith--Purcell emission of surface-plasmon polaritons in graphene}. We consider a geometry similar to Figure 2(a) but with an array of hBN disks instead of just an individual particle and setting $\epsilon = 1$ for simplicity [see inset in (b)]. \textbf{(a)}~Emission probability per particle $\Gamma_{\text{SP},n}(\omega)$ for different orders $n$ and fixed energy $\hbar\omega=174$~meV (SP wavelength $\lamp=299$~nm) as a function of $z_0$. We set the period to $a=2\,\lamp=598$~nm, while the rest of the disk and substrate parameters are the same as in Figure \ref{Fig2}. Emission angles $\theta_n$ for the propagating orders $n$ are indicated in the inset on a top view of the linear array (green circles). \textbf{(b)}~Emission probability per particle for orders shown in (a) as a function of period $a$ for $z_0=40$~nm and $v=0.04\,c$. The dashed-black line marks the value $a=598$ nm used in panel (a). Each order $n$ exhibits a vertical asymptote at a specific value of $a$ below it is not allowed. \textbf{(c)}~Emission probability per particle for $n=-4$ as a function of $a$ and $v/c$ for $z_0=40$~nm, with all other parameters the same as in (b). The white region corresponds to conditions for which $n=-4$ Smith--Purcell emission is kinematically forbidden. The conditions $\cos\theta_{-4}=-1$ and $\cos \theta_{-4}=0$ are indicated by dashed-blue and dashed-orange curves, respectively.}
\label{Fig4}
\end{figure*}

\subsection{Smith--Purcell emission with an array of hBN nanodisks}

In Figure~\ref{Fig4}, we consider a linear array of hBN nanodisks with similar substrate and size parameters as in Figure~\ref{Fig2}. For simplicity, we assume $\epsilon=1$ (this condition is nearly met by amorphous silica at the mid-infrared frequencies under consideration), so we can apply the above formalism and calculate the probability from Eq.~(\ref{Gammaarray}) with $p_x(\omega)=\alpha_{xx}^{\rm eff}(\omega)E_{\elec,x}(0,\omega)$ and $p_z(\omega)=0$. Here, we use Eqs.~(\ref{ErefandEdir}) for $E_{\elec,x}(0,\omega)$ with the reflection coefficient $r_{\kpar{\rm p}}$ taken from Eq.~(\ref{tilderp}); we set $\alpha_{xx}^{\rm eff}(\omega)=[\alpha_{xx}(\omega)-\mathcal{G}_{xx}(\omega)]^{-1}$ with $\alpha_{xx}(\omega)$ defined by Eq.~(\ref{eq_polarizability_disk}); and $\mathcal{G}_{xx}(\omega)$ is computed by numerical integration of Eq.~(\ref{Gsurffinal}) (see Sec.~\ref{latticesum} in Methods). We note that $\bar{\epsilon}=\mathcal{R}_\p=1$ because of the choice of $\epsilon=1$.

Figure~\ref{Fig4}(a) shows the Smith--Purcell emission probability for all allowed orders. Similar to Figure~\ref{Fig3}, the Smith--Purcell emission probability is maximized at an optimum value of $z_0$ that is rather independent of $n$. This effect can be understood for the disks under consideration because there is no out-of-plane polarization (i.e., $p_z=0$), and therefore, the dependences of Eq.~(\ref{Gammaarray}) on either $n$ or $z_0$ can be factorized. Interestingly, the emission is strongest for $n=-4$ because the chosen lattice period $a$ is close to the onset of this order [see Figure~\ref{Fig4}(b)], so the emission angle $\theta_{-4}=162\degree$ is close to grazing [see inset of Figure~\ref{Fig4}(a)] and the probability scales as $1/\sin\theta_{-4}$ [see Eq.~(\ref{Gammaarray})]. In Figure~\ref{Fig4}(c), we show the onset of the $n=-4$ order by a dashed-blue curve corresponding to the condition $\theta_{-4}=180\degree$, which makes $\Gamma_{{\rm SP},-4}$ divergent. Conversely, $\cos\theta_{-4}=90\degree$ and $\Gamma_{{\rm SP},-4}$ vanishes along the dashed-orange line [see also Eq.~(\ref{Gammaarray})].

\section{Conclusions}

In summary, we show that free electrons can efficiently excite SPs assisted by the mediation of scatterers placed in the vicinity of a polariton-supporting surface. Our semi-analytical theory shows that the emission probability is maximized for lossless resonant scatterers placed at an optimum distance from the surface. This method enables the emission of surface polaritons even if the electron cannot directly excite polaritons due to kinematic constraints. In particular, polaritons of relatively large phase velocity can still be excited by low-energy electrons. We further explore polaritonic Smith--Purcell emission mediated by linear arrays of scatterers under excitation by electrons moving parallel to the array. An optimum array--surface separation is also observed for maximum emission. Interestingly, lattice resonances emerge as singularities in the Smith--Purcell excitation probability, signaled by the onset of new emission orders when their corresponding polariton wavelengths match the period of the array.

To excite strongly confined surface modes, we are interested in small scatterers compared to the light wavelength. This condition is hard to reconcile with the requirement of lossless particles, but some suitable choices can be found depending on the spectral range of interest. For example, we show that hBN disks are excellent resonant scatterers at mid-infrared frequencies, where they can sustain photon-polaritons with reasonably low material losses that are overshadowed by those associated with their coupling to SPs. Our results show that these particles can assist electron coupling to graphene plasmons under conditions for which direct coupling is kinematically forbidden. Mie modes in dielectric particles offer another possibility in a spectral range for which materials with a relatively high refractive index are available, such as silicon and germanium in the near-infrared. For visible frequencies, noble-metal nanoparticles can feature plasmonic resonances that only depend on morphology in the quasistatic limit (particle size much smaller than the light wavelength), so they can be made arbitrarily small down to a few nanometers (when nonlocal effects become relevant), although intrinsic material losses in the material are substantial. We further envision arrays formed by optically trapped two-level atoms as an appealing approach to realize lossless resonant scatterers.

Unlike other methods of SP generation, the use of free electrons to excite SPs presents the advantage that no structuring of the surface is necessary. Through mediation of nanoparticles decorating the surface at an optimum distance, we show that SPs can be efficiently excited at positions determined by the position of the particles and under conditions for which direct electron--polariton coupling is not possible when using low-energy electrons. In addition, resonant particles allow spectral selectivity of the generated polaritons, filtering specific frequencies from the broadband field provided by the moving electrons. For particle arrays, a directional emission of SPs is predicted, displaying interesting features associated with lattice resonances that deserve further exploration in future studies.

\section{Methods}

\subsection{Optical response of the materials under consideration}
\label{appendix optical response}

The optical response of supported graphene is described through its surface conductivity in the Drude model [Eq.~(\ref{Drude})], with $\wD=\EF/\pi\hbar$ depending on the doping Fermi energy $\EF$. For silver, the permittivity can be well-approximated by a modified Drude model $\epsilon(\omega)=\epsilon_{\rm b}-\wbulk^2/\omega(\omega+\ii\gamma)$ with parameters $\epsilon_{\rm b}=4$, $\hbar\wbulk=9.17$~eV, and $\hbar\gamma=21$~meV extracted from optical measurements \cite{JC1972,paper300}; in our calculations, we set $\epsilon_{\rm b}=1$ for simplicity, as this parameter does not influence the results significantly for the plasmon frequencies under consideration. For thin hBN, polarization in the material is dominated by in-plane directions described by the corresponding permittivity component $\epsilon_{\rm hBN}(\omega)$, and consequently, we ignore the out-of-plane permittivity and approximate
\cite{CKC14}
$\epsilon_{\rm hBN}(\omega)=\epsilon_\infty-f\omega_0^2/[\omega(\omega+\ii \gamma)-\omega_0^2]$ with parameters
$\epsilon_\infty=4.90$, $f=2.00$, $\hbar\omega_0=168.6$~meV,
and $\hbar\gamma=0.87$~meV in the upper Reststrahlen band.

An effective surface conductivity $\sigma(\omega)=\ii\omega d[1-\epsilon(\omega)]/4\pi$ can be extracted from the material permittivity $\epsilon(\omega)$ for a film of small thickness $d$ compared with the polariton wavelength. We apply this approach to describe silver and hBN disks, by setting $\epsilon(\omega)$ to the respective permittivities of these materials (see above). For silver, this procedure leads to the Drude conductivity in Eq.~(\ref{Drude}) with $\wD=\hbar\wbulk^2 d/4\pi e^2$.

\subsection{Effective polarizability of a thin disk near an embedded 2D layer}
\label{appendix polarizability}

In the electrostatic limit, the in-plane polarizability of a thin disk of diameter $D$ supported on a homogeneous substrate of permittivity $\epsilon$ can be expressed as \cite{paper303}
\begin{subequations}
\label{alphadisk}
\begin{align}
\label{eq_polarizability_disk}
\alpha_\parallel(\omega)=\bar{\epsilon}\,D^3\sum_j\dfrac{\zeta_j^2}{1/\eta(\omega)-1/\eta_j}
\end{align}
in terms of contributions arising from different polaritonic eigenmodes $j$, which enter through their associated dipolar matrix elements $\zeta_j$ and eigenvalues $\eta_j$. Here, $\bar{\epsilon}=(\epsilon+1)/2$ is the average permittivity of the media on both sides of the disk, while $\eta(\omega)=\ii\sigma(\omega)/\bar{\epsilon}\omega D$ captures the response of the disk material through its surface conductivity $\sigma(\omega)$. We dismiss the out-of-plane polarizability (i.e., $\alpha_\perp=0$) under the assumption that the disk thickness $d$ is small compared with $D$. The response is dominated by the lowest-order dipolar mode $j=1$, for which one finds the parameters \cite{paper303} $\zeta_1(x) = a\ee^{bx}+c$ and $\eta_1(x)=a'\ee^{b'x}+c'$, which depend on the thickness-to-diameter ratio $x=d/D$ and is parametrized with material-independent constants $a=-0.01267$, $b=-45.34$, $c=0.8635$, $a'=0.03801$, $b'=-8.569$, and $c'=-0.1108$. For hBN disks, we set the thickness to $d=1$~nm (roughly 3 hexagonal atomic monolayers).

The polarizability in Eq.~(\ref{eq_polarizability_disk}) assumes a thick substrate of permittivity $\epsilon$. However, under the configuration considered in Figures~\ref{Fig2} and \ref{Fig4}, a 2D material layer [surface conductivity $\sigma(\omega)$] is placed at a distance $z_0$ below the outer surface on which the disk is lying. This layer changes the polarizability to an effective value $\alpha^{\rm eff}_\parallel(\omega)$ given by an expression analogous to Eq.~(\ref{alphaeff}), but with $\mathcal{G}_\parallel(\omega)$ reinterpreted as the field that is self-induced by the presence of the 2D layer on a unit parallel dipole placed at the position of the disk [i.e., the effect of the homogeneous substrate of permittivity $\epsilon$ is already included in Eq.~(\ref{eq_polarizability_disk})]. Considering the p-polarization electrostatic reflection coefficient $r'_{\kpar\p}=\kpar/(\kpar-\kp')$ with $\kp'=\ii\omega\epsilon/2\pi\sigma(\omega)$ for a 2D layer embedded in an infinite material of permittivity $\epsilon$ \cite{GP16} [cf. this expression and Eq.~(\ref{eq_pole_approx}) for a supported 2D layer], as well as $r_0=(\epsilon-1)/(\epsilon+1)$ for the substrate alone, we find
\begin{align}
\alpha^{\rm eff}_\parallel(\omega)=\frac{1}{1/\alpha_\parallel(\omega)-\tilde{\mathcal{G}}_\parallel(\omega)}
\label{alphadiskfinal}
\end{align}
with $\alpha_\parallel(\omega)$ given by Eq.~(\ref{eq_polarizability_disk}) and
\begin{align}
\tilde{\mathcal{G}}_\parallel(\omega)=\frac{2\epsilon}{(\epsilon+1)^2}\int_0^\infty\kpar^2\d\kpar\,\frac{r'_{\kpar\p}\ee^{-2\kpar z_0}}{1+r_0r'_{\kpar\p}\ee^{-2\kpar z_0}}.
\end{align}
\end{subequations}
Compared to Eq.~(\ref{Gqs}), this expression incorporates an overall factor $4\epsilon/(\epsilon+1)^2$ arising from the forward and backward field transmission across the outer planar surface, as well as a denominator accounting for multiple scattering in the cavity defined by the surface and the 2D material. We use Eq.~(\ref{alphadiskfinal}) for the effective polarizability of the disk in Figures~\ref{Fig2} and \ref{Fig4}.

\subsection{2D polariton emission probability by an induced dipole}
\label{polempro}

We can obtain the SP emission probability from the energy absorbed by the 2D material in the configuration of Figure~\ref{Fig1}(a) under the assumption that the response is dominated by SPs. The energy absorbed by a material of permittivity $\epsilon_{\rm m}(\omega)$ occupying a region $V$ and exposed to an optical field $\Eb(\rb,\omega)$ (in the frequency domain) is given by \cite{J99} $\int_0^\infty\d\omega\,\hbar\omega\,\Gamma(\omega)$, where
\begin{align}
\Gamma(\omega)=\frac{1}{4\pi^2\hbar}\int_V\d^3\rb\;{\rm Im}\{\epsilon_{\rm m}(\omega)\}\,|\Eb(\rb,\omega)|^2. \label{Gammagen}
\end{align}
For the 2D material, we consider an arbitrarily small thickness $d$, which yields a high permittivity $\epsilon_{\rm m}(\omega)=1+4\pi\ii\sigma(\omega)/\omega d$ expressed in terms of the surface conductivity $\sigma(\omega)$. The normal electric field inside the material is therefore negligible because of the continuity of the normal displacement, while the parallel field created by the dipole at the surface plane is continous and given by Eqs.~(\ref{dipolefields}) with $z=-z_0$. In the electrostatic limit, this field reduces to
\begin{align}
\Eb_{{\rm dip}\parallel}(\Rb,-z_0,\omega)=&
\frac{-\ii}{2\pi}\int\d^2\kparb \,\ee^{\ii \kparb\cdot\Rb-\kpar z_0} \label{field2Dfilm}\\
&\times (r_{\kpar\p}-1)\,\big[\pb(\omega)\cdot(\ii\hat{\kb}_\parallel+\zz)\big]\,\kparb. \nonumber
\end{align}
Inserting this result into Eq.~(\ref{Gammagen}), carrying out the $\Rb$ integral analytically, and reducing the $z$ integral to a factor $d$, we find
\begin{align}
\Gamma(\omega)=&\frac{1}{\hbar\omega}{\rm Re}\{\sigma(\omega)\} \nonumber\\
&\times\int_0^\infty\kpar^3\d\kpar \,\ee^{-2\kpar z_0}
|r_{\kpar\p}-1|^2 \big[|\pb_\parallel|^2+2|p_z|^2\big], \nonumber
\end{align}
which, upon consideration of Eq.~(\ref{eq_pole_approx}) and the dispersion relation $\kp=\ii\omega\bar{\epsilon}/2\pi\sigma(\omega)$ for the supported film in Figure~\ref{Fig1}(a), finally transforms into Eq.~(\ref{GSP}) with ${\rm Im}\{\mathcal{G}_\parallel(\omega)\}$ defined by Eq.~(\ref{Gqs}). To obtain this result, we have used the expressions ${\rm Re}\{\sigma(\omega)\}=(\omega\bar{\epsilon}/2\pi){\rm Im}\{-1/\kp\}$ and $\kpar|r_{\kpar\p}-1|^2=\mathcal{R}_\p{\rm Im}\{r_{\kpar\p}\}/{\rm Im}\{-1/\kp\}$ (with $\mathcal{R}_\p=1/\bar{\epsilon}$) derived from the relations between $\kp$, $\sigma(\omega)$, and $r_{\kpar\p}$.

For the SP emission probability in the configuration of Figure~\ref{Fig2}(a), the in-plane field acting on the 2D layer takes a similar form as Eq.~(\ref{field2Dfilm}), but now setting $p_z=0$ for the disk, replacing $r_{\kpar\p}$ by the reflection coefficient $r'_{\kpar\p}$ of the fully $\epsilon$-embedded 2D film (see Sec.~\ref{appendix polarizability}), and inserting an additional factor $t_0/(1+r_0r'_{\kpar\p}\ee^{-2\kpar z_0})$ in the integrand to account for the transmission of the dipole field across the topmost surface [transmission coefficient $t_0=2\sqrt{\epsilon}/(\epsilon+1)$] as well as multiple Fabry-Perot scattering within the surface--2D-film cavity. Then, by following the same steps as above but using the so-modified field, we trivially obtain Eq.~(\ref{GSPhBN}) with the Green tensor defined in Eq.~(\ref{GqshBN}).

\subsection{Dipole--dipole interaction in linear arrays}
\label{latticesum}

In a linear array, the dipole self-interaction needs to be supplemented by the contribution of the field induced by other dipoles. We thus redefine $\mathcal{G}(\omega)=\mathcal{G}^{\rm dir}(\omega)+\mathcal{G}^{\rm ref}(\omega)$ as the sum of a direct term contributed by all other dipoles in the array plus a surface-reflected term produced all dipoles. Starting from Eqs.~(\ref{dipolefields}) and adopting the electrostatic limit, the modified Green tensor is found to have vanishing off-diagonal components. The direct contribution can easily be evaluated by numerically summing the analytical expression for the free-space dipole--dipole interaction, yielding $\mathcal{G}_{xx}^{\rm dir}(\omega)=2S^{\rm dir}$ and $\mathcal{G}_{yy}^{\rm dir}(\omega)=\mathcal{G}_{zz}^{\rm dir}(\omega)=-S^{\rm dir}$ with
\begin{align}
S^{\rm dir}=\frac{2}{a^3}\sum_{j=1}^{\infty}\frac{\cos(j\omega a/v)}{j^3}. \nonumber
\end{align}
For the surface-reflected components, we follow a procedure similar to the derivation of Eq.~(\ref{phiarray1}) and write
\begin{align}
&\big[\mathcal{G}_{xx}^{\rm surf}(\omega),\mathcal{G}_{yy}^{\rm surf}(\omega),\mathcal{G}_{zz}^{\rm surf}(\omega)\big] \label{Gsurffinal}\\
&=\frac{1}{a}\sum_{n=-\infty}^\infty\int_{-\infty}^\infty\frac{\d k_y}{k_{n\parallel}}\,[k_{nx}^2,k_y^2,k_{n\parallel}^2]\,\ee^{-2k_{n\parallel}z_0}r_{k_{n\parallel}\p}, \nonumber
\end{align}
with $k_{nx}=\omega/v+2\pi n/a$ and $k_{n\parallel}=\sqrt{k_{nx}^2+k_y^2}$. Now, we consider low-loss SPs and a surface reflection coefficient satisfying Eq.~(\ref{Imrpll}), which allows us to evaluate the $k_y$ integral and write
\begin{align}
&{\rm Im}\big\{\big[\mathcal{G}_{xx}^{\rm surf}(\omega),\mathcal{G}_{yy}^{\rm surf}(\omega),\mathcal{G}_{zz}^{\rm surf}(\omega)\big]\big\} \label{latticesumfinal}\\
&\approx\frac{2\pi\mathcal{R}_\p}{a}\sum_{n=-\infty}^\infty\frac{\kp}{\sqrt{\kp^2-k_{nx}^2}}\,[k_{nx}^2,\kp^2-k_{nx}^2,\kp^2]\,\ee^{-2\kp z_0}, \nonumber
\end{align}
for the surface-reflected component of the Green tensor in the array, where $\mathcal{R}_\p=1/\bar{\epsilon}$ and $\kp$ are approximated as real numbers and the $n$ sum is restricted by $|k_{nx}|<\kp$ (see Supplementary Figure~\ref{FigS5}).

\noindent{\bf Research funding:} This work was supported by the European Research Council (Advanced grant 101141220-QUEFES), the European Commission (Horizon 2020 grant nos. FET-Proactive 101017720-eBEAM and FET-Open 964591-SMART-electron), the Spanish MCINN (PID2020-112625GB-I00 and Severo Ochoa CEX2019-000910-S), the Catalan CERCA Program, and Fundaci\'os Cellex and Mir-Puig.


%

\clearpage
\pagebreak \onecolumngrid \section*{Supplementary Figures}
\renewcommand{\thefigure}{S\arabic{figure}}

\begin{figure*}[htbp]
\centering
\includegraphics[width=0.7\linewidth]{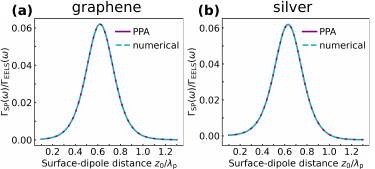}\caption{\textbf{Electron coupling to surface polaritons mediated by a small scatterer: numerical vs analytical approaches}. Probability of polariton emission $\Gamma_{\rm SP}(\omega)$ normalized to the EELS probability $\Gamma_{\rm EELS}(\omega)$ as a function of scatterer--surface distance $z_0$ for \textbf{(a)} graphene (emission energy $\hbar\omega=120$\,meV, Fermi energy $\EF=1~{\rm eV}$) and \textbf{(b)} silver ($\hbar\omega=1.75$\,eV). The probability is calculated either numerically (dashed-light-blue curves) or analytically through the polariton-pole approximation (PPA, purple curves). In (b), we consider a $1$~nm-thick film. We set $v=0.04c$ and $k_{\rm D}b=0.53$ in both panels.}
\label{FigS1}
\end{figure*}

\begin{figure*}[htbp]
\centering
\includegraphics[width=0.7\linewidth]{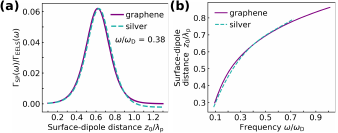}
\caption{\textbf{Electron coupling to surface polaritons mediated by a small scatterer: graphene vs silver}. \textbf{(a)}~Probability of surface-plasmon emission $\Gamma_{\rm SP}(\omega)$ normalized to the EELS probability $\Gamma_{\rm EELS}(\omega)$ for a graphene monolayer (Fermi energy $\EF=1~{\rm eV}$) and a silver film (thickness $d=1~{\rm nm}$) as a function of the scattered--surface distance $z_0$ normalized to the corresponding plasmon wavelength $\lambda_{\rm p}$ for a fixed emission frequency $\omega=0.38\,\omega_{\rm D}$. \textbf{(b)}~Optimum surface--dipole distance that maximizes $\Gamma_{\rm SP}(\omega)/\Gamma_{\rm EELS}(\omega)$ as a function of frequency for the same materials as in (a). We set $v=0.04c$ and $k_{\rm D}b=0.53$ in both panels.} 
\label{FigS2}
\end{figure*}

\begin{figure*}[htbp]
\centering
\includegraphics[width=0.7\textwidth]{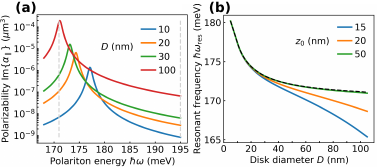}
\caption{\textbf{Size-dependence of hBN disk polarizabilities.} {\bf (a)}~Imaginary part of the in-plane polarizability of hBN disks of 1\,nm thickness and different diameters $D$ (see labels) supported on a homogeneous substrate of permittivity $\epsilon=2$, as calculated from Eq.~(25a) in the main text. {\bf (b)}~Peak energy extracted from (a) as a function of disk size (dashed-black curve), compared with the peak energy obtained from the effective disk polarizability under the configuration of Fig.~2a [Eq.~(25b)] for different separations $z_0$ (see labels). We set the graphene Fermi energy to $\EF=1~{\rm eV}$ and the hBN disk thickness to $d=1\,{\rm nm}$.}
\label{FigS3}
\end{figure*}

\begin{figure*}[htbp]
\centering
\includegraphics[width=0.7\textwidth]{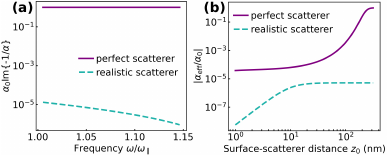}
\caption{\textbf{Effective polarizability of an hBN disk and a perfect (lossless and resonant) dipole}. \textbf{(a)}~Frequency dependence of the imaginary part of the inverse polarizabilities $\Imm\{-1/\alpha\}$ for a perfectly scattering dipole [$\alpha=\ii\alpha_0$ with $\alpha_0=3\epsilon/2k_1^3$ and $\epsilon=2$] and an hBN disk [Eq.~(25a) in the main text], normalized to $1/\alpha$. The frequency is normalized to the in-plane resonant frequency of hBN, $\omega_\parallel=171.1\,{\rm meV}$. \textbf{(b)} Absolute value of the in-plane component of the effective polarizability $\alpha_{\rm eff}$ given by Eq.~(25b) in the main text and normalized to $\alpha_0$ for the same two scatterers as in (a), but placed at a distance $z_0$ from a graphene layer (Fermi energy $\EF=1~{\rm eV}$). We present results as a function of $z_0$ for fixed energy $\hbar\omega=180$~meV. The hBN disk thickness and diameter are $d=1\,{\rm nm}$ and $D=20\,{\rm nm}$ in all cases.}
\label{FigS4}
\end{figure*}

\begin{figure*}[htbp]
\centering
\includegraphics[width=1.0\textwidth]{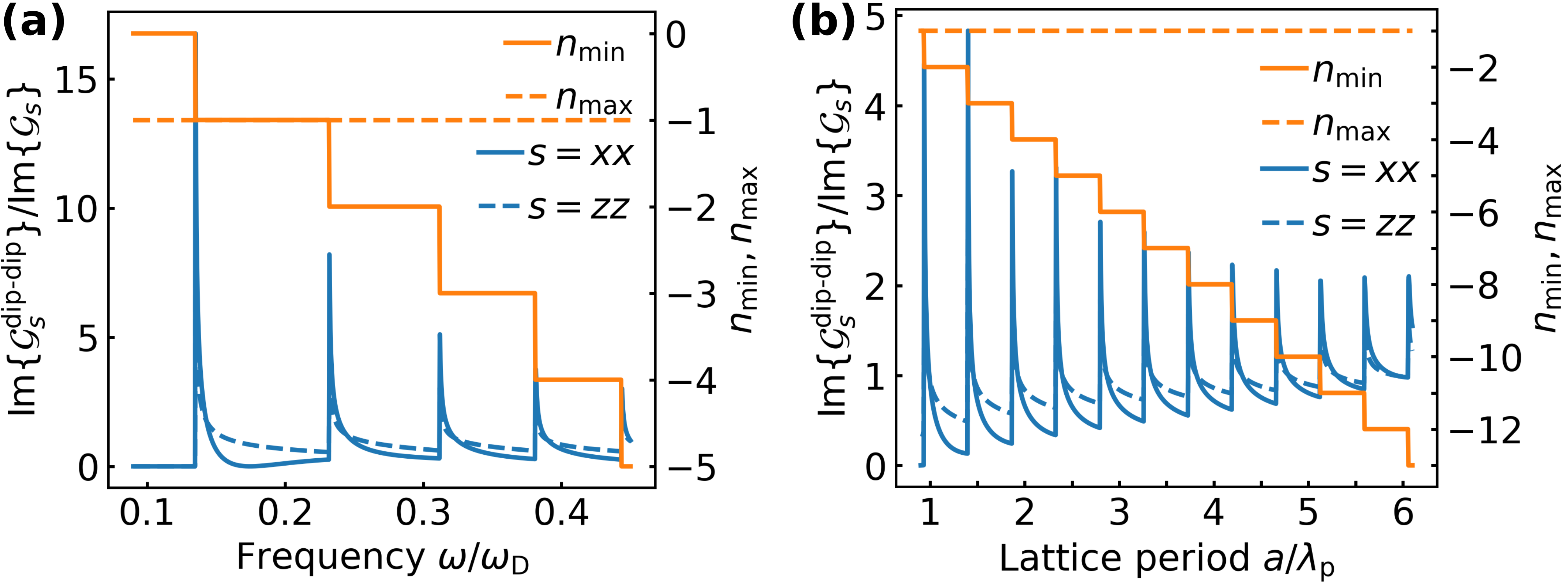}
\caption{\textbf{Lattice resonances in the dipole--dipole interaction of linear periodic arrays}. \textbf{(a)}~Frequency dependence of the ratio ${\rm Im}\{\mathcal{G}^{\rm surf}_{s}\}/{\rm Im}\{\mathcal{G}^{\rm particle}_{s}\}$ between the Green tensor components of a particle array [only the surface contribution ${\rm Im}\{\mathcal{G}^{\rm surf}_{s}\}$, calculated from Eq.~(29) in the main text] and a single particle [${\rm Im}\{\mathcal{G}^{\rm particle}_{s}\}$, taken from Eq.~(12)] for $s=xx$ (equivalent to $\parallel$ in the single particle) and $s=zz$ ($\perp$ in the single particle). The frequency is normalized to $\omega_{\rm D}$. In the array, the ratio of the period to the polariton wavelength is set to $a/\lamp=2$. \textbf{(b)}~Same ratio of Green-tensor components as in panel (a), plotted as a function of $a/\lamp$ for a fixed frequency $\omega=0.4\,\omega_{\rm D}$. We use the parameters $z_0=0.6\,\lamp$, $k_{\rm D}b=0.1$, and $v=2.2v_0$ in both panels. The right axes indicate the upper and lower limits of the sum in Eq.~(29): $n_{\rm max}$ (dashed-orange curve) and $n_{\rm min}$ (solid-orange curve), respectively.}
\label{FigS5}
\end{figure*}

\end{document}